\documentstyle[12pt]{article}
\newlength{\extraspace}
\newlength{\extraspaces}
\newcommand{\be}{\begin{equation}
\addtolength{\abovedisplayskip}{\extraspaces}
\addtolength{\belowdisplayskip}{\extraspaces}
\addtolength{\abovedisplayshortskip}{\extraspace}
\addtolength{\belowdisplayshortskip}{\extraspace}}
\newcommand{\ee}{\end{equation}}
\newcommand{\ba}{\begin{eqnarray}
\addtolength{\abovedisplayskip}{\extraspaces}
\addtolength{\belowdisplayskip}{\extraspaces}
\addtolength{\abovedisplayshortskip}{\extraspace}
\addtolength{\belowdisplayshortskip}{\extraspace}}
\newcommand{\ea}{\end{eqnarray}}
\newcommand{\nonu}{\nonumber \\[.5mm]}

\newcommand{\e}{\, {\rm e}}
\newcommand{\D}{{\cal D}}
\newcommand{\bra}[1]{\left\langle {#1} \right\vert}
\newcommand{\ket}[1]{\left\vert {#1} \right\rangle}
\newcommand{\VEV}[1]{\left\langle {#1} \right\rangle}

\newcommand{\A}{&\!\!\!}

\begin{document}
\addtolength{\baselineskip}{.7mm}
\thispagestyle{empty}
\begin{flushright}
TIT/HEP--205 \\
August, 1992
\end{flushright}
\vspace{13mm}
\begin{center}
{\large{\bf{ $c \! = \! 1$ TWO DIMENSIONAL QUANTUM GRAVITY}}}
\footnote{A talk given at the International Symposium on
Quantum Physics and the Universe,
Waseda University, August 19-22, 1992} \\[18mm]
{\sc  Norisuke Sakai} \\[10mm]
{\it Department of Physics, Tokyo Institute of Technology \\[2mm]
Oh-okayama, Meguro, Tokyo 152, Japan} \\[10mm]
{\bf Abstract}\\[7mm]
{\parbox{13cm}{\hspace{5mm}
The continuum (Liouville) approach to the
two-dimensional (2-D) quantum gravity is reviewed
with particular attention to the $c=1$ conformal matter coupling,
and new results on a related problem of dilaton gravity
are reported.
After finding the physical states,
we examine the procedure to compute correlation functions.
The physical states in the relative cohomology
show up as intermediate state poles of the correlation functions.
The states in the absolute cohomology but not in the relative cohomology
arise as auxiliary fields in string field theory.
The Liouville approach is applied also to the quantum treatment of
the dilaton gravity.
The physical states are obtained from the BRST cohomology and
correlation functions are computed in the dilaton gravity.
}}
\end{center}
\vfill
\newpage
\vspace{13mm}
\begin{center}
{\large{\bf{ $c \! = \! 1$ TWO DIMENSIONAL QUANTUM GRAVITY}}}
 \\[15mm]
{\sc  Norisuke Sakai} \\[10mm]
{\it Department of Physics, Tokyo Institute of Technology \\[2mm]
Oh-okayama, Meguro, Tokyo 152, Japan} \\[10mm]
{\bf Abstract}\\[7mm]
{\parbox{13cm}{\hspace{5mm}
The continuum (Liouville) approach to the
two-dimensional (2-D) quantum gravity is reviewed
with particular attention to the $c=1$ conformal matter coupling,
and new results on a related problem of dilaton gravity
are reported.
After finding the physical states,
we examine the procedure to compute correlation functions.
The physical states in the relative cohomology
show up as intermediate state poles of the correlation functions.
The states in the absolute cohomology but not in the relative cohomology
arise as auxiliary fields in string field theory.
The Liouville approach is applied also to the quantum treatment of
the dilaton gravity.
The physical states are obtained from the BRST cohomology and
correlation functions are computed in the dilaton gravity.
}}
\end{center}
%\vfill
%\newpage
%
\setcounter{section}{0}
\setcounter{equation}{0}
%
%%%%%%%%%%  Introduction  %%%%%%%%%%%%%%%%%%%%%%%%%%%%%%%%%%%%%%%%%
%
\section {Introduction}
%\numberbysection
In recent years, advance in the matrix model \cite{BIPZ}, \cite{GRMI}
for the nonperturbative treatment of two-dimensional
(2-D) quantum gravity has stimulated a significant progress in the
continuum approach by means of the Liouville theory.
There are two main motivations to study the 2-D quantum
gravity coupled to matter.
{}First, it is precisely a string theory when the 2-D
space-time is regarded as the world sheet for the string.
Second, it provides a toy model for the quantum gravity in higher
dimensions such as four dimensions.
\par
The matrix model is a discretized approach to the 2-D quantum gravity
and is very powerful in providing a nonperturbative treatment.
It also gives a precise definition of quantum gravity since
the theory is regularized from the beginning.
On the other hand, the Liouville theory is a continuum approach
\cite{DDK}, \cite{SEIREV} and is complementary
to the discretized approach of the matrix model in many respects.
{}First, the physical meaning is more transparent in the continuum
approach.
{}For instance, the identification of
physical states is easier in the Liouville theory.
The method of BRST cohomology is particularly useful in this respect.
In spite of the nonlinear dynamics of
the Liouville theory, a method based on conformal field theory has
now been sufficiently developed to
understand the results of the matrix model
and to offer a powerful method in computing
various quantities.
In particular, we can calculate not only partition
functions but also correlation
functions by using the procedure of an analytic
continuation (in terms of external momenta and/or the central charge
$c$) \cite{GOULI}, \cite{DFKU}.
Second,  the continuum approach is more flexible to incorporate
various related issues, although one has to use certain Ansatz
and an appropriate regularization
to deal with the Liouville theory.
{}For instance, supersymmetric model for the 2 dimensional gravity
is available so far only in the continuum approach \cite{DFKU},
 \cite{AODH}.
As another such example, we will briefly report our results on the
quantum theory of the dilaton gravity in 2 dimensions that are
discussed much recently with the black hole evaporation
 \cite{CGHS}--\cite{MASATAUC}.
\par
So far only conformal field theories with central charge $ c \le 1$
have been successfully coupled to quantum gravity.
The $c = 1$ model is the richest and the most interesting, and it
is in some sense the most easily soluble \cite{GMIL}.
On the other hand, the $c = 1$ model is a borderline case for the
present treatment of the 2-D quantum gravity, and therefore it might
give some clues to go beyond the $c =1$ barrier.
{}From the viewpoint of the string theory, the $c=1$ model
has at least one (continuous) dimension of target space in which
strings are embedded.
Therefore, we can discuss the space-time interpretation in the usual
sense in the $c=1$ model.
Since the Liouville (conformal) mode plays a dynamical role if the
dimension of the target space is different from the usual critical
dimensions, the theory is called ``noncritical'' string theory.
\par
It has been observed that the $c=1$ quantum gravity can be regarded
effectively as a critical string theory in two
dimensions, since the Liouville field zero
mode provides an additional ``time-like'' dimension besides the
obvious single spatial dimension given by the zero mode of the $c=1$
matter \cite{POLCH}.
Namely the system can effectively be regarded as a (critical) string
theory in two-dimensional target space.
Since the coordinate on the world sheet is introduced entirely
for convenience, the model is invariant under reparametrizations
on the world sheet (two-dimensional diffeomorphisms) as usual in string
theory.
Because of the two-dimensional diffeomorphism invariance,
the oscillation in two tangent directions of the world sheet becomes
unphysical, and only the transverse oscillation remains physical.
Since we are effectively dealing with a type of two-dimensional
(critical) string,
there are no transverse oscillations left.
Thus we expect naively that there is only the center of mass motion of
the string, which lead to a physical scalar particle upon
quantization.
Though the resulting physical particle is massless, it is still
referred to as ``tachyon'' following
the usual terminology borrowed from the critical string theory.
Since there are no transverse directions, the continuous (field)
degrees of freedom are exhausted by the tachyon field.
The partition function for the torus topology
was computed in the Liouville theory, and was found to give
precisely the same partition function as the tachyon field
alone \cite{SATAPART}.
\par
However, it has been noted that there exist other
discrete degrees of freedom in the $ c=1$ matter coupled to the 2-D
quantum gravity \cite{GRKLNE}--\cite{SATAFACT}.
These states are called discrete states, or are sometimes called special
states or topological states.
It has been demonstrated that these discrete states show up as
intermediate state poles in the correlation functions.
They are the remaining quantum mechanical degree of freedom
of the string, and in a sense they may be regarded as remnants of
characteristic features of string theory such as winding modes.
These states fall into representations of $SU(2)$ and are associated
with generators of the area preserving diffeomorphisms
 \cite{WITTEN}.
The three point interactions among discrete states \cite{KLPO},
those among tachyons and discrete states \cite{MASATA}, and
those with nonvanishing ghost number states \cite{OHSU}
have been worked out by using the operator product expansion.
\par
The complete analysis of physical states has been obtained by a
BRST analysis \cite{LIAN}.
The usual discrete states are among the so-called relative cohomology
states that are obtained by restricting to the subspace built on
only one of ghost zero mode vacuum.
Without restricting to such a subspace, we have the so-called absolute
cohomology states.
We find recently that these states in the absolute cohomology but not
in the relative cohomology serve as auxiliary fields in the string
field theory \cite{SATASFT}.
\par
Recently many groups have studied the dilaton gravity in 2 dimensions
to discuss the black hole evaporation
\cite{CGHS}--\cite{MASATAUC}.
Most of the works have eventually employed the semi-classical
approximation, which is often blamed to be the
possible origin of diseases in this problem.
Therefore it is very desirable to have a full quantum treatment of the
dilaton gravity even for a restricted class of models.
We take the recently proposed models of dilaton
gravity \cite{BICA} that are conformally invariant and attempted
to study the quantum theory applying the methods used in the
Liouville theory.
We performed the BRST analysis in the case
of the number of matter fields $N\geq24$.
Apart from the usual string states with momentum and oscillator
excitations, we found that there are only a few
physical states like discrete states. This situation is very similar
to the case of $0\leq N<24$ analyzed in \cite{BILAL}, \cite{LIAN}.
We have also computed the correlation functions by applying the methods
used in the Liouville theory \cite{MASATAUC}.
An insertions of these gauge invariant operators
corresponds to creating a hole with appropriate boundary conditions
in the language of the physics on the world sheet.
If our analysis is extended to macroscopic
loops from the local operators, we can find quantum transitions
leading to topology change in the two-dimensional dilaton gravity.
In order to discuss the black hole evaporation, we need to prepare
the appropriate wave function for the black hole and to examine its
evolution.
We hope that our results on the physical states and
the correlation functions
may serve as a first step for a full quantum
treatment of the dilaton gravity beyond the usual semi-classical
approximations. Let us emphasize that the continuum approach is
flexible enough to study the dilaton gravity, whereas the powerful
method of matrix models is yet to be applied to this case.
However, while we are writing this paper, we received an interesting
preprint which computed the partition function with the tachyon
background for the $c=1$ quantum gravity by using matrix model
approach \cite{DIMOPL}. We are currently studying to
apply their results to our case of $N=0$ dilaton gravity.
The wave functions of the universe and possible consequences on the
black hole evaporation are being studied.
\par
%
%%%%%%%%%%  2-D Gravity as a Conformal Field Theory  %%%%%%%%%%%%%%%%%%%
%%%%%%%
%
\section { 2-D Gravity as a Conformal Field Theory }
We consider the c=1 conformal matter realized by a single
bosonic field $X$ coupled to the 2-D quantum gravity.
Since the Einstein action is a topological invariant in
two-dimensions, we have only a cosmological term as a gravitational
action
\be
S_{cosm} [g]
 = {\mu_0 \over \pi} \int_{M_h} d^2 z \sqrt{g}, \qquad
{1 \over 4\pi} \int_{M_h} d^2 z \sqrt{g} R
= 2-2h
\label{bcosm}
\ee
where a manifold with $h$ handles and
the bare cosmological constant are denoted as $M_h$ and $\mu_0$
respectively. Matter part is given by a usual free scalar action
\be
S_{matter} [g, X]
 = {1 \over 8\pi} \int_{M_h} d^2 z \sqrt{ g}
 g^{\alpha\beta} \partial_{\alpha} X \partial_{\beta} X .
\label{bmatter}
\ee
The partition function $Z$ is given as a sum over the partition function
$Z_h$ for a Riemann surface $M_h$ with the genus $h$.
The contribution $Z_h$ from the manifold with $h$ handles
corresponds to a quantum correction in $h$ loop order and is
given by a path integral
\be
Z= \sum_{h=0}^{\infty} Z_h, \qquad
Z_h=g_{st}^{2h-2} \int {{\cal D} g_{\alpha \beta}
{\cal D} X \over V_{\rm diffeo}}
{\rm e}^{- S_{matter}[g, X] - S_{cosm}[g]},
\label{partfunc}
\ee
where $g_{st}$ is the string coupling constant and $V_{\rm diffeo}$ is
the volume of the group of diffeomorphisms.
\par
A quantization of string theories in dimensions other than the critical
ones is considered in the conformal gauge
in refs. \cite{DDK} $g_{\alpha \beta}
= {\rm e}^{\phi_0} \hat g_{\alpha \beta}(\tau)$,
where $\phi_0$ is the Liouville field and $\hat g_{\alpha \beta}(\tau)$
is the reference metric that depends on the moduli parameters
$\tau$ of the Riemann surface.
We can convert the integration over the metric to the integration over
$\xi$, $\tau$, and $\phi_0$  \cite{POLCHINSKICMP} where $\xi(z)$ is the
infinitesimal diffeomorphism.
The change of integration variables leads to a Jacobian, whose
effect can be reproduced by a functional integral over ghost
$c$ and antighost $b$ with a ghost action
$S_{ghost}$.
Therefore the functional integral over the metric can be converted into
an integral over the ghosts, the Liouville field, and the moduli
\be
{{\cal D} g_{\alpha \beta} \over V_{\rm diffeo}}
= {d\tau \over V_{\rm CKV}}
{\cal D}_g \phi_0 {\cal D}_g b {\cal D}_g c
{\rm e}^{- S_{ghost}[g, b, c]},
\label{metricint}
\ee
where the volume $V_{\rm CKV}$ generated by conformal Killing vectors
remains.
\par
The dependence on the Liouville field in the functional measure for
the matter field and the ghost field can be given by a conformal
anomaly  \cite{FRIEDAN}
\be
{\cal D}_{{\rm e}^{\phi_0}\hat g} X
{\cal D}_{{\rm e}^{\phi_0}\hat g} b
{\cal D}_{{\rm e}^{\phi_0}\hat g} c
={\cal D}_{\hat g} X
{\cal D}_{\hat g} b
{\cal D}_{\hat g} c
{\rm e}^{-S_{anom}[\phi_0, \hat g]}.
\label{confanom}
\ee
The functional measure for the Liouville field depends on the Liouville
field itself and is much more difficult to deal with. David and Distler
and Kawai have proposed that one can transform this nonlinear measure
for the Liouville field into the usual translationally invariant
functional
measure for the free field which uses the reference metric rather
than the original metric \cite{DDK}
\be
 {\cal D}_{{\rm e}^{\phi_0}\hat g} \phi_0
={\cal D}_{\hat g} \phi_0 {\rm e}^{-J[\phi_0, \hat g]}.
\label{lioumeas}
\ee
{}From symmetry arguments, they assumed that the sum of the Jacobian $J$
and the anomaly terms should be the same functional form as the usual
Liouville type local action.
If we rescale the Liouville field
$\phi_0 = \alpha \phi $,
the cosmological term
should be proportional to $\sqrt{\hat g} {\rm e}^{\alpha \phi}$.
{}Following \cite{DDK}, we will choose the proportionality factor
$\alpha$ so that the kinetic term of the Liouville action
acquires the standard normalization for a scalar field
\be
J+S_{anom}+S_{cosm}=S_L[\phi, \hat g]
 \equiv  {1 \over 8\pi} \int d^2 z \sqrt{\hat g}
\Bigl( \hat g^{\alpha\beta} \partial_\alpha \phi \partial_\beta \phi
- Q \hat R \phi + 8 \mu \e^{\alpha \phi} \Bigr),
\label{liouville}
\ee
where the renormalized cosmological constant is denoted as $\mu$
The parameters $Q$ and $\alpha$
are determined by demanding that the theory should be
independent of the choice of the reference metric $\hat g$.
This condition implies the vanishing total conformal anomaly and
that the cosmological constant is an operator of the conformal
weight $(1, 1)$.
\be
Q=\sqrt{(25-c_{matter}) \over 3} \rightarrow 2\sqrt2, \quad
\alpha = -{Q \over 2} \pm \sqrt{Q^2 - 8 \over 2} \rightarrow
 -\sqrt2 \qquad c_{matter} \rightarrow 1.
\label{qvalue}
\ee
\par
%
%%%%%%%%%%  Physical States %%%%%%%%%%%%%%%%%%%
%%%%%%%
%
\section { Physical States }
The conformal technique is most powerful to discuss the case of
vanishing cosmological constant $\mu=0$, since the action
(\ref{liouville}) simply reduces to a free boson field
theory. We will consider mostly this case except for the case
where the Liouville zero mode needs to be treated more carefully
when we compute correlation functions.
The holomorphic part of the energy-momentum tensor of
the matter and the Liouville field is given for $c=1$ case by
\be
T_{X\phi}(z)
= -{1 \over 2} (\partial X)^2 -{1 \over 2} (\partial \phi)^2
- \sqrt2 \partial^2\phi.
\label{holemtensor}
\ee
In order to find the physical states in terms of these fields,
one has to look for primary fields $\Psi (z)$ of unit weight with
respect to this $T(z)$.
The open string vertex operators are given by line integrals of
primary fields with boundary conformal weight one along the
boundary, while the closed string vertex operators are given by
surface integrals of primary fields with conformal weight $(1, 1)$.
There are two types of physical operators:
tachyons and the discrete states.
Both types are present in open string theory and closed string theory.
\par
Let us first consider the holomorphic part only.
The simplest field for operators with unit conformal weight
is the tachyon vertex operator with momentum $p$
\be
\Psi^{(\pm)}_p (z) = \e^{ipX(z)} \e^{\beta^{\pm}\phi(z)},\qquad
\beta^{(\pm)}=-\sqrt2\pm p.
\label{tachyon}
\ee
The positive (negative) sign solution for the Liouville energy
$\beta^{+}$ ($\beta^{-}$) is usually called as positive (negative)
chirality.
This is the momentum eigenstate without the oscillator excitations.
{}For the level\footnote{
The level is the sum of the index of oscillator excitations.}
$n>0$ there are non-trivial primary fields
only when the momentum is an integer or a half odd
integer in unit of $\sqrt2$.
There are two solutions for the Liouville energy similarly to the
tachyon case
\be
\beta^{(\pm)}=-\sqrt2\pm \sqrt{p^2+2n}, \qquad p\in {\bf Z}/2.
\label{discreteenergy}
\ee
Seiberg has argued that only the solution with $\beta^{+}$ corresponds
to a local field. We call this case as Seiberg type, and the other
solution with $\beta^{-}$ as anti-Seiberg type.
They
are called
the ``discrete states'' \cite{GRKLNE}, \cite{POLYAKOV}
which form $SU(2)$ multiplets and can be constructed
as vertex operators \cite{WITTEN}, \cite{KLPO} with appropriate
cocycle factors \cite{MASATA}.
\par
The above analysis gives the physical states with the matter and
the Liouville field only.
The complete list of the physical states including the ghost sector
can be achieved by considering the BRST cohomology \cite{LIAN}.
\be
Q_{B} =\oint{dz \over 2\pi i}\Bigl[c(z)
\bigl(T_{X\phi}(z)+{1 \over 2}T_{gh}(z)\bigr)
+{3 \over 2}\partial^2c(z)\bigr]
=c_0L_0-b_0M+ \hat d
\label{brstcharge}
\ee
The physical states are defined as the states annihilated by the BRST
charge, modulo BRST transform of some states
\be
Q_{B} \bigl\vert \, {\rm phys} \bigr\rangle =0, \qquad
\bigl\vert \, {\rm phys} \bigr\rangle \not=
Q_{B} \bigl\vert \, \bigr\rangle =0.
\label{abscoh}
\ee
The above cohomology is called the absolute cohomology, since it is
considered in the entire space.
On the other hand, it is always useful to decompose $Q_B$ in term of
the ghost zero mode $c_0$ and $b_0$ and consider the cohomology in the
subspace restricted by an additional condition
space $b_0 \bigl\vert \, \bigr\rangle =0$.
In this subspace, one can use the restricted nilpotent operator $\hat d$
defined in eq.(\ref{brstcharge}) instead of the full BRST
operator $Q_B$.
\par
Both cohomologies have been worked out in ref.\cite{LIAN}.
{}For the level zero states (without the oscillator excitations),
relative as well as absolute cohomology consists of usual tachyons only.
On the other hand, the cohomology classes for the first level or higher
consist of discrete states and its partners.
{}For the Seiberg-type states, relative cohomology classes
consist of the discrete state and an accompanying state with ghost
number $-1$.
{}For instance at the first level,
\be
p^{\mu}\equiv(p,-i\beta)=(0,0), \qquad
\alpha_{-1} \bigl\vert \, 0 \bigr\rangle, \qquad
b_{-1} \bigl\vert \, 0 \bigr\rangle.
\label{seibergrel}
\ee
The remaining states in the absolute cohomology classes are
obtained from the states in the relative cohomology by multiplying
$c_0$ and an addition of certain states.
{}For instance at the first level,
\be
c_0\alpha_{-1} \bigl\vert \, 0 \bigr\rangle, \qquad
c_0b_{-1} \bigl\vert \, 0 \bigr\rangle
+{i \over \sqrt2}\phi_{-1} \bigl\vert \, 0 \bigr\rangle
\label{seibergabs}
\ee
{}For the anti-Seiberg-type states, relative cohomology classes
consist of the discrete state and an accompanying state with ghost
number $-1$. For instance at the first level,
\be
p^{\mu}=(0,-i(-2\sqrt2)), \qquad
\alpha_{-1} \bigl\vert \, p^{\mu} \bigr\rangle, \qquad
c_{-1} \bigl\vert \, p^{\mu} \bigr\rangle.
\label{antiseibrel}
\ee
Anti-Seiberg states in the absolute cohomology but not in the
relative cohomology at the first level are,
\be
p^{\mu}=(0,-i(-2\sqrt2)), \qquad
c_0\alpha_{-1} \bigl\vert \, p^{\mu} \bigr\rangle, \qquad
c_0c_{-1} \bigl\vert \, p^{\mu} \bigr\rangle.
\label{antiseibabs}
\ee
All the cohomology classes come as quartets similar to the first level
examples.
%
%
%%%% Correlation Functions and the Role of the Physical States %%%%%%%%%
%
\section { Correlation Functions and the Role of the Physical States}
In recent years, the method of the free boson conformal field theory
has been used to obtain correlation functions.
It has been argued that the method is reliable to compute the most
singular part of the amplitudes as we let $\mu \rightarrow 0$
\cite{POLYAKOV}, \cite{DFKU}.
\par
One of the obvious physical state is the tachyon with appropriate
gravitational dressing
\be
O_p = \int d^2z \Psi_p(z)
 = \int d^2z \sqrt{\hat g} \; {\rm e}^{ipX} \,
      {\rm e}^{\beta(p)\phi} , \qquad
\beta^{(\pm)}(p) = -\sqrt2 \pm p.
\label{tachyonop}
\ee
Let us consider correlation functions of the gravitationally
dressed tachyon vertex operators
\ba
<\prod_{j=1}^n {O_{p_j}}>
&\!\!\! = &\!\!\!
\int {d\tau  \over V_{CKV}}{\cal D} \phi {\cal D} X   \;\,
     {\rm e}^{- S[ X, \phi, b, c, \hat g]} \;\,
     O_{p_1} \cdots O_{p_n}  \nonu
&\!\!\! = &\!\!\!
 \int \prod_{i=1}^n  d^2 z_i \, {d\tau \over V_{CKV}}
     <{{\rm e}^{ip_1 X(z_1)} \cdots {\rm e}^{ip_n X(z_n)}}>_X
 <{{\rm e}^{\beta_1 \phi(z_1)} \cdots
     {\rm e}^{\beta_n \phi(z_n)}}>_\phi,
\label{corxphi}
\ea
The integration over the zero mode of $X$ field just gives momentum
conservation, but the zero mode $\phi_0$ for the Liouville field
is of the form
\ba
\int_{-\infty}^{\infty} d\phi_0 \,
{\rm e}^{-\alpha s\phi_0-B{\rm e}^{\alpha \phi_0}}
&\!\!\! = &\!\!\!
   {1 \over \alpha} \, \Gamma(-s) \, B^s  \nonu
s= {1 \over -\alpha} \left( Q(1-h) + \sum_{i=1}^n\beta_{i} \right),
& & \qquad
B(\tilde \phi)={\mu \over \pi}\int d^2z \sqrt{\hat g}
{\rm e}^{\alpha \tilde \phi},
\label{phizeroint}
\ea
where we have assumed that $s<0$ and $B(\tilde \phi)>0$ in order for the
integration to converge.
The integral is effectively cut-off by a term
$B{\rm e}^{\alpha \phi_0}$ in the action for large negative
values of $\phi_0$, whereas it is cut-off by
a term $\alpha s\phi_0$ for large positive values of $\phi_0$
 \cite{POLCH}.
The former case corresponds to the long distance limit and the latter
to the short distance limit, since the original metric is given by
$g_{\alpha, \beta}=\hat g_{\alpha, \beta}{\rm e}^{\alpha \phi} $
with a negative $\alpha$.
The above procedure shows that the zero mode integration is
limited both at the ultraviolet and at the infrared.
If we are interested in dominant (singular) contributions to
amplitudes in the limit
$\mu \rightarrow 0$, we can safely ignore
the cosmological term except for the fact that the zero mode integration
region is limited to a region of length proportional to ln$\mu$.
Such dominant (singular) part of the amplitude is called bulk amplitude
or resonant amplitudes \cite{DFKU}, \cite{POLYAKOV}.
\par
Thus the zero mode integration gives the $s$-th power of the
cosmological term with the nonzero mode.
\ba
 & & \qquad
<\prod_{j=1}^n O_{p_j}>
=  2\pi \delta \left( \sum_{j=1}^N p_j \right)
    {1 \over -\alpha} \Gamma(-s) \tilde A (p_1, \cdots, p_N), \nonu
 \tilde A &\!\!\! = &\!\!\!
    \int \prod_{i=1}^N
    {d\tau \over V_{CKV}}
 d^2 z_i \sqrt{\hat g}
  < \prod_{j=1}^N {\rm e}^{ip_j {\tilde X}(z_j)}>_{\tilde X}\,
    < \left( {\mu \over \pi} \int d^2 w \,
    \sqrt{\hat g} \, {\rm e}^{\alpha \tilde \phi(w)} \right)^s
    \prod_{j=1}^N {\rm e}^{\beta_j \tilde \phi(z_j)}>_{\tilde\phi}.
\label{kntotal}
\ea
By analytically continuing to
nonnegative integer values of $s$ (an analytic continuation
in external momenta and/or the central charge),
we can evaluate the non-zero mode integral by means of the usual
free field contractions.
\par
It has been shown that the $N$-tachyon amplitude with the chiralities
$(-,+,\cdots,+)$ is given for the sphere topology by
\cite{GOULI}, \cite{DFKU}
\be
\tilde A (p_1, \cdots, p_N)
=  {\pi^{N-3} [\mu\Delta (-\rho)]^s \over \Gamma(N+s-2)}
   \prod_{j=2}^N \Delta(1 - \sqrt{2} p_j),
\label{nptfin}
\ee
where  $\Delta (x)=\Gamma(x)/ \Gamma(1-x)$.
The regularization parameter $\rho$ is given by $\rho = - \alpha^2 /2$
and is eventually set equal to $ -1$. The seemingly divergent quantity
$\mu\Delta (-\rho)$ should be interpreted \cite{GOULI}, \cite{DFKU}
as the renormalized cosmological constant $\mu_r$.
\par
It has been observed that the poles in the correlation functions can be
explained as the short-distance singularities between tachyons with
opposite chirality \cite{SATAFACT}, \cite{DFKU}.
Namely, the pole at $p_j=1/\sqrt2$ is found to come from the tachyon
intermediate state, whereas the poles at
$p_j=(n+1)/\sqrt2, \, n\in{\bf Z}$ are due to the discrete states.
One may think of singularities due to intermediate states in other
channels such as the short-distance singularities between the negative
chirality tachyon and a number of positive chirality tachyons.
The residues of these poles are found to involve
amplitudes with two or more particles for each type $+$ or $-$.
It has been shown that amplitude with only one particle for one of
the types $+$ or $-$ is nonvanishing, but the other configurations
vanish in generic momenta \cite{DFKU}, \cite{SATAFACT}.
Thus we found that all the poles in the correlation functions can
be understood as the short-distance singularities and that the
factorization (unitarity at the tree level, i.e., on the sphere
topology)
is valid despite the peculiar structure of the
correlation functions (\ref{nptfin}).
\par
By factorizing the correlation functions, one can reach three point
couplings among discrete states.
However, such a coupling is most easily obtained by using the operator
product expansion \cite{KLPO}.
By constructing the appropriate cocycle factors, we have obtained
all possible three point couplings among discrete states and tachyon
completely  \cite{MASATA}. Three point couplings involving states with
nonvanishing ghost number are computed also in ref. \cite{OHSU}.
These three point couplings materialize the underlying symmetry
called area-preserving diffeomorphisms
and summarized in algebraic terms as the ground ring
\cite{WITTEN}--\cite{KLPO}.
\par
Since the states that appeared as the intermediate states are
in the relative BRST cohomology, it is interesting to seek for the
role played by the states in the absolute cohomology but not in the
relative cohomology.
{}For that purpose, we have constructed the kinetic term of
string field theories and examined the component
decomposition \cite{SATASFT}
\ba
S_{\rm open}
&\!\!\! = &\!\!\!  \sum_{n=-2}^{2} \int d^2 x
\e^{i Q \cdot x} \bra{\Phi_{-n}(x)} Q_B \ket{\Phi_n(x)},  \nonu
&\!\!\! = &\!\!\!  \int d^2 x \e^{i Q \cdot x} \biggl[ \,
{1 \over 2} \partial_\mu \phi \partial_\mu \phi - \phi^2
+ {1 \over 4} ( \partial_\mu A_\nu - \partial_\nu A_\mu )^2 \nonu
&\!\!\! &\!\!\!  + 2 (\psi - {1 \over 2} (\partial + i Q) \cdot A)^2
- (\partial_\mu \psi^\ast + 2 A_\mu^\ast) \partial_\mu \chi
+ \cdots \, \biggr].
\label{opencompaction}
\ea
where $\phi$ denotes the tachyon, $A_{\mu}$ discrete state,
$\psi^\ast, A_{\mu}^\ast, \chi$
states in the relative cohomology, whereas $\psi$ denotes
the state in the absolute cohomology but not in the relative
cohomology.
We find that the field $\psi$ is an auxiliary field that has no
kinetic term.
In the case of the closed string, one needs to use the so-called
semi-relative cohomology instead of the absolute cohomology
\cite{WIZW}
\be
(Q_{B}+\bar Q_B) \bigl\vert \, {\rm phys} \bigr\rangle =0, \quad
(b_0-\bar b_0) \bigl\vert \, {\rm phys} \bigr\rangle =0, \quad
\bigl\vert \, {\rm phys} \bigr\rangle \not=
(Q_{B}+Q_{B}) \bigl\vert \, \bigr\rangle =0.
\label{abscohcl}
\ee
We constructed the component decomposition also for the closed string
theory and found again that
the state in the absolute cohomology but not in the relative
cohomology is an auxiliary field that has no kinetic
term \cite{SATASFT}.
\par
%
%
%%%% Dilaton-Gravity in 2-Dimensions %%%%%%%%%
%
\section { Dilaton-Gravity in Two Dimensions }
Recently there has been a tremendous activity in studying the
two-dimensional gravity interacting with a dilaton and matter
fields since the work on the black hole evaporation \cite{CGHS}.
Although the Einstein action is a topological invariant in
two-dimensions, it acquires a dynamical meaning when multiplied by a
scalar field
\ba
S_{dilaton}
& =&  {1 \over 2\pi} \int d^2 z \sqrt{\bar g}
\bigl[-{\rm e}^{-2\phi} \bar R +2\mu\bigr] \nonu
& = & {1 \over 2\pi} \int d^2 z \sqrt{g}{\rm e}^{-2\phi}\bigl[- R
-4(\nabla\phi)^2+2\mu\bigr].
\label{dilatonaction}
\ea
where we used the metric with Euclid signature and have performed a
conformal transformation $g_{\mu\nu}={\rm e}^{2\phi}\bar g_{\mu\nu}$
in the latter form.
Let us note that the fields multiplying the Einstein action can be
absorbed into the metric except in two dimensions.
The latter form of the dilaton gravity system is suggested by the
string theory and has been discussed in connection
with black hole evaporation.
There have been extensive analyses of the Hawking radiation
and its back reaction using semi-classical approximation.
It is often noted that there may be a disease or inadequacy in the
semi-classical approximation.
In two dimensions, we can perhaps hope to understand the dilaton
gravity quantum mechanically without using the semi-classical
approximation.
We attempt to use the techniques learnt in the Liouville approach
to two-dimensional gravity in this dilaton gravity.
There appeared recently an interesting approach using the
conformal field theory to discuss the problem quantum mechanically
\cite{BICA}.
\par
We use the conformal gauge $g_{\mu\nu}={\rm e}^{2\rho}\hat g_{\mu\nu}$
with the Liouville field $\rho$ and the reference metric
$\hat g_{\mu\nu}$.
Changing the path integral measure for the matter fields to the
translation invariant measure using $\hat g_{\mu\nu}$, we obtain
the conformal anomaly
\ba
\D_{\e^{2\rho} \hat g} f \D_{\e^{2\rho} \hat g} b
\D_{\e^{2\rho} \hat g} c
\A = \A \D_{\hat g} f \D_{\hat g} b \D_{\hat g} c \,
\e^{- S_{\rm anomaly}^{{\rm gh}, f}}, \quad
S^{{\rm gh}, f}_{\rm anomaly}  =  {26-N \over 12}
S_{\rm L}[\rho, \hat g], \nonu
S_{\rm L}[\rho, \hat g]
\A = \A {1 \over 2\pi} \int d^2 z \sqrt{\hat g}
\left( \hat g^{\mu\nu} \partial_\mu \rho \partial_\nu \rho
+ \hat R \rho \right),
\label{anomalymatter}
\ea
Since the measure for the Liouville field is nonlinear,
the Jacobian for the transformation to the translation invariant
measure using $\hat g_{\mu\nu}$ is more difficult to determine.
It has been argued that this Jacobian must be a local functional
of the Liouville field $\rho$ containing a bilinear kinetic term,
a term linear in the curvature and $\rho$, and the exponential
term representing the cosmological term \cite{DDK}.
This is an extremely successful Ansatz.
The quantization of the dilaton is similar but may be more complicated.
There has also been a proposal for the use of metric
${\rm e}^{-2\phi}g_{\mu \nu}$ rather than $g_{\mu \nu}$ in defining
the ghost measure \cite{CGHS}. Therefore, along the spirit of
ref.\cite{DDK}, we take the following more
general Ansatz for the Jacobian for the transformation of the
functional measure to the translationally invariant measure using
$\hat g_{\mu\nu}$ in the case of the dilaton gravity with
the matter
\ba
S \A = \A S_{\rm kin} + S_{\rm cosm}, \qquad
S_{\rm kin}  =  S_{\rm dilaton}(\mu=0) + S_{\rm anomaly}
+ S_{\rm matter} \nonu
S_{\rm kin}
\A = \A {1 \over 2\pi} \int d^2 z \sqrt{\hat g}
\biggl[ {\rm e}^{-2\phi} \left(
-4 \hat g^{\mu\nu} \partial_\mu \phi \partial_\nu \phi
+ 4 \hat g^{\mu\nu} \partial_\mu \phi \partial_\nu \rho
- \hat R \right) \nonu
\A\A - \kappa \left( \hat g^{\mu\nu} \partial_\mu \rho
\partial_\nu \rho + \hat R \rho \right)
+ a \left(2 \hat g^{\mu\nu} \partial_\mu \phi \partial_\nu \rho
+ \hat R \phi \right) \nonu
\A\A + b \hat g^{\mu\nu} \partial_\mu \phi \partial_\nu \phi
+ {1 \over 4} \sum_{j=1}^N
\hat g^{\mu\nu} \partial_\mu f^j \partial_\nu f^j \biggr].
\label{kinetic}
\ea
This is the most general Ansatz for the anomaly under the assumption
that the measures are defined by ${\rm e}^{\alpha\phi}g_{\mu\nu}$
for the quantization of various fields.
This is a slight generalization of the
Ansatz in ref.\ \cite{BICA}. The cosmological term $S_{\rm cosm}$
will be determined after we discuss the physical states.
\par
We can reduce the above kinetic term to a free field action by a
change of variables.
In the case of $\kappa\not=0$, the change of variables reads
\ba
\omega \A = \A {\rm e}^{-\phi}, \quad
\chi = -{\rho \over 2}-{\omega^2 + a\ln\omega \over 2\kappa}
={\hat \chi \over 4\sqrt{|\kappa|}}, \nonu
d\Omega \A = \A -{1 \over \kappa}
d\omega \sqrt{\omega^2 - \kappa + a
+ {a^2 + b\kappa \over 4\omega^2}}
={d\hat \Omega \over 4\sqrt{|\kappa|}}.
\label{fieldredef}
\ea
This change of variables (\ref{fieldredef})
gives a free field action with the source term for $\hat \chi$
\be
S_{\rm kin} = {1 \over 8\pi} \int d^2 z \sqrt{\hat g} \biggl( \mp
\hat g^{\mu\nu} \partial_\mu \hat \chi \partial_\nu \hat \chi
\pm 2 \sqrt{|\kappa|} \hat R \hat \chi
\pm \hat g^{\mu\nu} \partial_\mu \hat \Omega
\partial_\nu \hat \Omega + \sum_{j=1}^N
\hat g^{\mu\nu} \partial_\mu f^j \partial_\nu f^j \biggr).
\label{dilkin}
\ee
The free field theory is almost identical to the usual Liouville
theory with matter fields, except that
there is one field with negative metric.
The transformed Liouville field $\hat \chi$ has negative metric
in the case of $\kappa>0$, whereas the transformed dilatonic
field $\hat \Omega$ has negative metric in the case of $\kappa<0$.
For $\kappa = 0$, the appropriate change of variables is
given by
\be
\chi^\pm = Q \left(- \rho - \frac{b}{2a} \phi
- \frac{2a+b}{4a} \ln \left| \e^{-2\phi} + \frac{a}{2} \right|
\right) \pm \frac{2}{Q} \left( \e^{-2\phi} - a\phi \right).
\label{fieldredefzero}
\ee
In this case, $\chi^+ + \chi^-$ transforms as the
Liouville field, while $\chi^+ - \chi^-$ transforms as a scalar
field. The free field action reads
\ba
S_{\rm kin} \A = \A {1 \over 8\pi} \int d^2 z \sqrt{\hat g} \,
\biggl( \hat g^{\mu\nu} \partial_\mu \chi^+ \partial_\nu \chi^+
- Q \hat R \chi^+ \nonu
\A\A - \hat g^{\mu\nu} \partial_\mu \chi^-
\partial_\nu \chi^- + Q \hat R \chi^- + \sum_{j=1}^N
\hat g^{\mu\nu} \partial_\mu f^j \partial_\nu f^j \biggr).
\label{dilkinzero}
\ea
Actions with different values of $Q$ are related by a
${\rm O}(1, 1)$ transformation in $\chi^+, \chi^-$ field space and
are equivalent. Finally the parameter $\kappa$ is determined by
requiring the conformal invariance (independence on
$\hat g_{\mu\nu}$) of the quantum theory (\ref{dilkin}) or
(\ref{dilkinzero}) \cite{BICA}
\be
\kappa={N-24 \over 12}.
\label{kappa}
\ee
\par
Let us note that the translation invariant path integral measure
for these free fields defines the quantization of the dilaton
gravity. In terms of the original variables, the path integral
measure may be quite nonlinear, and the path integral region
(values of the fields) may sometimes be peculiar.
Our
attitude is that the dilaton gravity is nonlinear in the
original variable and is difficult to quantize it.
Therefore we can use the free field representation
of the kinetic term (\ref{dilkin}) together
with the translation invariant measure as the definition of (the
kinetic term of) the quantum theory of the dilaton gravity.
\par
%
%%%%%%%  BRST Analysis  %%%%%%%%%%%%%%%%%%%%%%%%%%%%%%%%%%%%%%%%
%
In order to discuss the physical states and the gauge invariant
operators of the theory, we compute the cohomology of the BRST
charge $Q_{\rm B}$.
For the case of $0<N<24$, the analysis of ref.\ \cite{BILAL}
applies with $c=N+1$.
Since we have not found the results for $N \ge 24$ in
the existing literature, we compute the BRST cohomology for
$N \ge 24$ as a direct extension of other cases
\cite{LIAN}, \cite{BILAL}, and find a result
similar to the case $0<N<24$.
Let us first describe the relative cohomology states.
{}For $(p^{\Omega},p^j)\not=0$, the BRST cohomology
consists of
only the usual string excitations with vanishing ghost number,
\ba
{\rm dim} H^n_{rel}&=&\delta_{n,0}P_D(R), \qquad R\in {\bf Z}, \nonu
\Pi_{m=1}^{\infty}(1-q^m)^{-D}&=&\sum_{m=0}^{\infty}q^m P_D(m), \quad
{1 \over 2} p^I p^J \eta_{IJ} - i \sqrt{\kappa} p_{\chi} + R = 1.
\label{stringexcit}
\ea
where we have defined a metric for the free fields appropriate
to the $N>24$ case: $\eta_{IJ}=(-1,+1,\cdots,+1)$ for the fields
$(\chi, \Omega, f^j)$.
Even if $(p^{\Omega},p^j)=0$, the result is the same unless $p_I=0$
or $(p_{\chi},p_{\Omega},p_j)=(-2i\sqrt{\kappa}, 0,\cdots, 0)$.
Nontrivial cohomology similar to the discrete states in the $c=1$
gravity appear only for the following few cases:
\begin{enumerate}
\item For the $p_I = 0$,
we have $N+1$ states with vanishing ghost number
and one state with ghost number $-1$
\be
\alpha_{-1}^{\Omega} \ket{p_I=0}, \quad
\alpha_{-1}^{j} \ket{p_I=0}, \quad b_{-1} \ket{p_I=0},
\label{zeromom}
\ee
where $\ket{p_I}$ is the state with momentum $p_I$ annihilated by
all positive mode oscillators as well as $b_0$.
\item For
$p_I\equiv(p_{\chi},p_{\Omega},p_j)=(-2i\sqrt{\kappa}, 0, 0)$,
we have $N+1$ states with vanishing ghost number
and one state with ghost number $+1$
\be
\alpha_{-1}^{\Omega} \ket{p_I}, \quad \alpha_{-1}^{j} \ket{p_I},
\quad c_{-1} \ket{p_I}.
\label{twomom}
\ee
\end{enumerate}
The remaining states in the absolute cohomology is obtained by
multiplying by $c_0$ and by adding certain terms if necessary.
We have also worked out the semi-relative cohomology in the case of
closed string in the same way as in \cite{WIZW}.
We have also worked out the cohomology for $N=24$
and found a similar result \cite{MASATAUC}.
\par
The cosmological term is chosen from these BRST cohomology
classes by imposing the additional requirement \cite{BICA} that it
should be of the same form as the original cosmological term in the
limit of weak gravitational coupling ${\rm e}^{\phi}\rightarrow 0$.
We obtain uniquely
\be
S_{cosm}=\left\{
\begin{array}{ll}
{\mu \over \pi} \int d^2 z \sqrt{\hat g}
\e^{{1 \over \sqrt{|\kappa|}}(-\hat \chi+\hat \Omega)}
& \qquad \mbox{for $\kappa \neq 0$} \\
{\mu \over \pi} \int d^2 z \sqrt{\hat g}
\e^{-{1 \over Q}(\chi^+ + \chi^-)}
& \qquad \mbox{for $\kappa = 0$}
\end{array} \right..
\label{cosmdil}
\ee
The particularly interesting is the case of no additional matter
fields, i.e.,\ $N=0$ ($\kappa=-2$).
By identifying $\kappa=-2$ in eq.\ (\ref{dilkin}),
we find that the dilaton gravity system without additional matter
corresponds to the case of the Liouville gravity coupled to
the $c=1$ conformal matter. The only distinction is that the $c=1$
matter comes from the dilaton degree of freedom, and that the
resulting free boson $\hat \Omega$ has negative metric.
To compare with the physical operators in the Liouville gravity,
we should rotate the free boson to purely imaginary values
$\hat \Omega=i\bar \Omega$ and identify it with the usual free
boson with the positive metric. Then we find that the two-momentum
of this cosmological term is precisely the simplest discrete value
\cite{WITTEN} $(\beta_{\chi},p_{\bar\Omega})
=(-ip_{\chi},ip_{\Omega})=(-1/\sqrt{2}, 1/\sqrt{2})$.
\par
Among these BRST cohomplogy classes, we take
momentum eigenstates
(gravitationally dressed tachyon vertex operators in the
language of the Liouville gravity), and
have computed their correlation functions on the sphere \cite{MASATAUC}.
In the case of $N \neq 24$
\ba
O_p = \int d^2z \sqrt{\hat g} \, \e^{\beta_{\chi} \hat \chi
+ \beta_{\Omega}\hat \Omega+ip_j f^j}, \A\A \nonu
-{1 \over 2} \beta_{\chi}(\beta^{\chi}+2\sqrt{|\kappa|})
-{1 \over 2} \beta_{\Omega}\beta^{\Omega}+{1 \over 2}p_j p^j \A = \A 1.
\label{gaugeinvop}
\ea
{}First we perform the integration over the zero mode of the
Liouville-like field $\hat \chi$ and then integrate over the zero
mode of $\tilde \Omega$.
In this way we obtain the path integral with a number of cosmological
constant insertions whose momenta is kept at a generic value $f$
for a regularization.
The number of cosmological term insertions
$s$ is given by
\be
s =  \sqrt{|\kappa|}  \sum_{k=1}^n \beta_{\chi k}
- 2 \kappa(1-h).
\label{defofs}
\ee
\par
For the case of $N \not= 24, 0$, we find the
three-point correlation function on the sphere
\ba
\VEV{\prod_{k=1}^n {O_{p_k}}}
\A = \A (2\pi)^{N+1} \delta \left( \sum_{k=1}^n p_k \right)
     \delta \left( \sum_{k=1}^n \beta_{\Omega k}
     + {s \over \sqrt{|\kappa|}} \right)
     \sqrt{|\kappa|} \, \Gamma(-s)
     \tilde A (p_1, \cdots, p_n), \nonu
\tilde A (p_1, p_2, p_3)
\A=\A \bigl[ \mu \Delta(1 + p_1 \cdot q) \Delta(1 + p_2 \cdot q)
\Delta(1 + p_3 \cdot q) \bigr]^s,
\label{threepoint}
\ea
where $\Delta(x) \equiv \Gamma(x) / \Gamma(1-x)$.
For the $N=24$ case we obtain very similar three-point correlation
functions on the sphere \cite{MASATAUC}.
\par
In the case of $N=0$, we can obtain correlation functions for
arbitrary numbers $n$ of gauge invariant operators
(\ref{gaugeinvop}). Since the cosmological term in this case is at the
special discrete momentum, the correlation functions becomes
singular due to their insertions \cite{DFKU}.
Therefore we need to keep the momentum $q_I$ of the cosmological term at
a generic value to regularize the correlation functions.
After an analytic continuation in $s$ analogous to the
$c=1$ Liouville gravity theory, we obtain the $n$-point correlation
functions.
In the case of $N=0$, it is important to distinguish two solutions of
the on-shell condition (\ref{gaugeinvop})
\be
\beta_{\chi}^{(\pm)}=-\sqrt{2}\pm\beta_{\Omega}
\label{chirality}
\ee
Following the terminology of Liouville gravity \cite{},
we call the solution with the $+$ $(-)$ sign as positive (negative)
chirality.
We find the correlation functions of $n$ gauge invariant operators with
chiralities $(-,+, \cdots, +)$
\ba
\VEV{\prod_{k=1}^n O_{p_k}} \A=\A
2\pi \delta \left( {1 \over \sqrt2}\sum_{k=1}^n
(\beta_{\chi k} + \beta_{\Omega k}) + 2
\right) \Gamma(-s) \nonu
\A\A \times \left[\mu\Delta (-\rho) \right]^s
{\pi^{n-3}  \over \Gamma(n+s-2)}
\prod_{k=2}^n \Delta(1 - \sqrt{2} \beta_{\Omega k}),
\label{dilcor}
\ea
where the regularization parameter $\rho = q \cdot q / 2$ should
be set to zero eventually.
There is a striking similarity between the correlation function of
dilaton gravity and that of the Liouville gravity (\ref{kntotal}),
(\ref{nptfin}).
In a similar but somewhat different context,
the dilaton gravity correlation
function is also singular due to the discrete mometum for the
cosmological term.
We can absorb the divergence as a renormalization of the cosmological
constant
\be
\mu_r=\mu \Delta(-\rho).
\label{rencosm}
\ee
Correlation functions vanish identically for chirality
configurations other than the above one. When computing these
correlation functions, one should keep the regularization
parameter $\rho$ to be nonvanishing in order to avoid ambiguous
results.
\par
\par
\vspace{5mm}
{\large{\bf Acknowledgements}}
\begin{small}
\par
We thank Y. Tanii, Y.Matsumura, and T. Uchino for the
collaboration.
We also thank H. Kawai, T. Yoneya, T. Mishima, A. Hosoya,
and A. Nakamichi for a useful discussion.
This work is supported in part by
Grant-in-Aide for Scientific Research for Priority Areas
from the Ministry of Education, Science and Culture (No. 04245211).
\vspace{5mm}
\end{small}
\end{document}